# Technological Excellence Requires Human and Social Context

Karl Palmås,[1*] Mats Benner,[2] Monica Billger,[3] Ben Clarke,[4] Raimund Feifel,[5]
Julia Fernandez-Rodriguez,[6,7] Anna Foka,[8] Juliette Griffié,[9] Claes Gustafsson,[10]
Kerstin Hamilton,[11] Johan Holmén,[12] Kristina Lindström,[13,14] Tobias Olofsson,[15]
Joana B. Pereira,[16] Marisa Ponti,[4] Julia Ravanis,[17] Sviatlana Shashkova,[5]
Emma Sparr,[18] Pontus Strimling,[19,20] Fredrik Höök,[17,18,21*] Giovanni Volpe[5,22*]

1. Division of Science, Technology and Society, Chalmers University of Technology, 412 96 Gothenburg, Sweden.
2. Department of Business Administration, Lund University, 223 63 Lund, Sweden.
3. InfraVis – National Research Infrastructure for Data Visualization, Department of Architecture and Civil Engineering, Chalmers University of Technology, 412 96 Gothenburg, Sweden.
4. Department of Applied Information Technology, University of Gothenburg, 405 30 Gothenburg, Sweden.
5. Department of Physics, University of Gothenburg, 412 96 Gothenburg, Sweden.
6. Centre for Cellular imaging, Sahlgrenska Academy, University of Gothenburg, 405 30 Gothenburg, Sweden.
7. Integrated Microscopy Technologies, Science for Life Laboratory, University of Gothenburg, 405 30 Gothenburg, Sweden.
8. Department of ALM, Uppsala University, 753 13 Uppsala, Sweden.
9. Department of Biochemistry and Biophysics, Stockholm University, Stockholm, Sweden.
10. Institute of Biomedicine, University of Gothenburg, 405 30 Gothenburg, Sweden.
11. HDK-Valand – Academy of Art and Design, University of Gothenburg, 405 30 Gothenburg, Sweden.
12. Department of Engineering Science, University West, 461 86 Trollhättan, Sweden.
13. Malmö Research Centre for Imagining and Co-Creating Future, Malmö University, 205 06 Malmö, Sweden.
14. The School of Arts and Communication, Malmö University, 205 06 Malmö, Sweden.
15. Stockholm Center for Organisational Research (SCORE), Stockholm University, 106 91 Stockholm, Sweden.
16. Neuro Division, Department of Clinical Neurosciences, Karolinska Institute, Stockholm.
17. Department of Physics, Chalmers University of Technology, 412 96 Gothenburg, Sweden.
18. Department of Chemistry, Lund University, 221 00 Lund, Sweden.
19. The Institute for Futures Studies, 101 31 Stockholm, Sweden.
20. The Institute for Analytical Sociology, Linköping University, 601 74 Norrköping, Sweden.
21. Science for Life Laboratory, Chalmers University of Technology, 412 96 Gothenburg, Sweden.
22. Science for Life Laboratory, Department of Physics, University of Gothenburg, 412 96 Gothenburg, Sweden.

**Breakthrough technologies increasingly shape social institutions, economic systems, and political futures. Yet models of research excellence associated with such technologies often prioritize technical performance, scalability, and short-term innovation metrics while treating ethical, social, and cultural dimensions as secondary considerations. This perspective article argues that such separation is no longer tenable. Technological excellence requires sustained integration of human and social context—not as external oversight, but as a constitutive element of research quality. We propose a broader understanding of excellence that combines technical rigor with ethical robustness, social intelligibility, and long-term relevance. The rapid emergence of generative and agentic artificial intelligence further underscores this argument. As technological systems increasingly operate through language, interpretation, and normative alignment, expertise traditionally cultivated in the humanities and social sciences becomes integral to the design, governance, and responsible deployment of such systems. Drawing on historical examples and contemporary research practices, this article examines five interconnected domains**

where the humanities and social sciences, treated as integrated dimensions of research practice, can strengthen technological development: (1) ethical, legal, and social integration in agenda-setting and research design; (2) plural and reflexive foresight practices that shape technological futures; (3) graduate education as a leverage point for cross-disciplinary literacy; (4) visualization and communication as epistemic and civic practices; and (5) institutional frameworks that move beyond rigid distinctions between basic and applied research. Across these dimensions, we propose practical strategies for embedding interdisciplinary collaboration structurally rather than symbolically. Our aim is not to prescribe a single institutional model, but to articulate principles that support interdisciplinary research systems capable of producing technologies that are not only innovative and transformative, but also socially robust and meaningfully embedded in the contexts they transform.

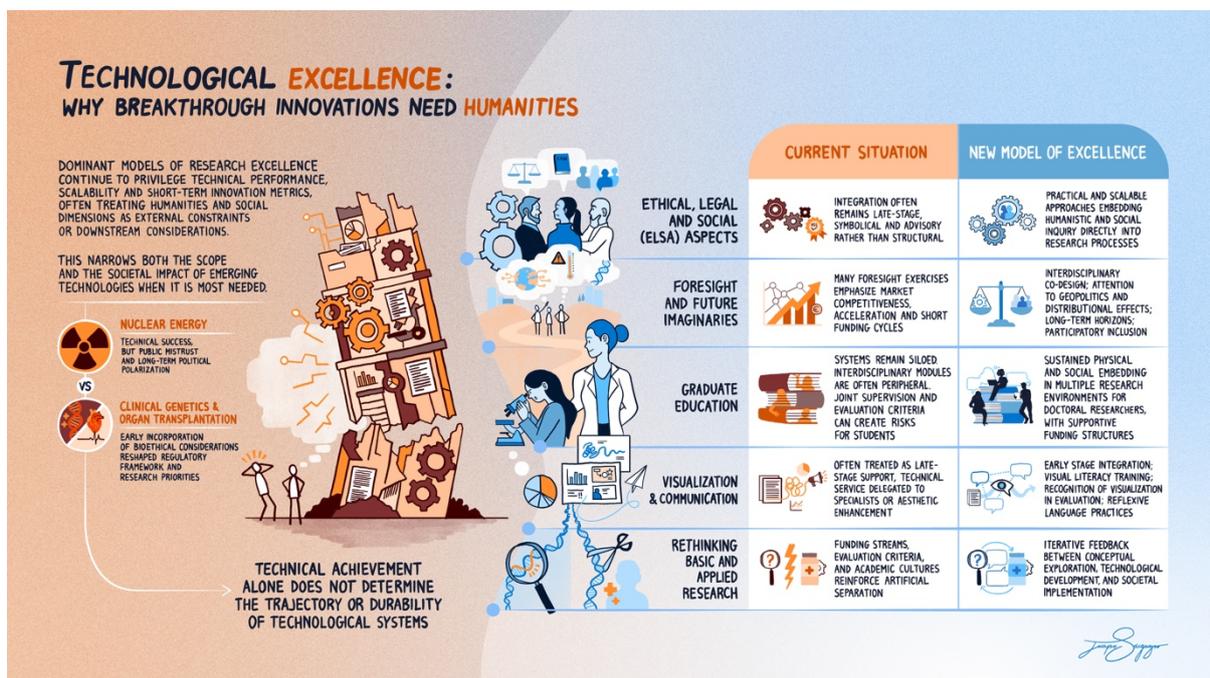

**Figure 1. Why breakthrough research needs humanities and social sciences.** Artwork: Jacopo Sacquegno.

**TECHNOLOGICAL EXCELLENCE REQUIRES HUMAN AND SOCIAL CONTEXT**     **1**

# Introduction

Breakthrough technologies increasingly shape how societies understand themselves, organize collective life, and imagine possible futures. Yet their transformative power depends not only on technical performance, but on the human expertise, institutional frameworks, and social environments in which they are embedded. From artificial intelligence and advanced materials to biotechnologies and climate-related innovations, technological research today unfolds in direct entanglement with social values, institutional structures, ethical commitments, and political choices. Yet dominant models of research excellence continue to privilege technical performance, scalability, and short-term innovation metrics, often treating human and social dimensions as external constraints or downstream considerations. This separation risks narrowing both the scope and the impact of technological research at precisely the moment when broader integration is most needed.

The history of technological innovation provides clear evidence of this dynamic. The development of nuclear energy, for example, was driven by extraordinary advances in physics and engineering, yet insufficient early integration of ethical reflection and democratic governance contributed to public mistrust and long-term political polarization.[1] By contrast, the early institutionalization of bioethical frameworks in clinical genetics and organ transplantation reshaped regulatory architectures and research priorities, facilitating broader societal legitimacy.[2] Similar dynamics characterize the evolution of information and communication technologies. Social media platforms and AI-driven decision systems were initially developed with strong emphasis on technical scalability, predictive accuracy, and market expansion. However, limited early integration of safeguards concerning misinformation, algorithmic bias, privacy, and democratic accountability contributed to widespread societal concern and subsequent regulatory intervention.[3] These developments underscore that technological performance alone does not secure public trust or sustainable adoption. Recent advances in generative artificial intelligence further intensify this dynamic. Large language models and emerging agentic AI systems interact with users primarily through natural language, narrative framing, and interpretive context. As a result, linguistic nuance, cultural assumptions, and normative judgments become operational elements of technological systems themselves. This development makes humanities and social sciences expertise relevant not only for evaluating technological impacts but also for shaping how such systems are designed, aligned, and governed.

Experience across multiple scientific domains shows that technological excellence cannot be sustained by technical ingenuity alone. Decisions about what problems are worth solving, which futures are desirable, and how risks and benefits are weighted and distributed are not peripheral to innovation but constitutive of it. At this level of problem framing and normative orientation, the humanities and social sciences play a decisive role: by clarifying values and assumptions, situating technologies within historical trajectories, interrogating power relations and governance structures, and expanding the imaginative horizons within which technological development takes place. Aside from such interpretive approaches, which seek to understand society and culture through description, the social sciences may also contribute with methods such as quantitative modelling and survey-based prediction, which seek to explain socio-cultural mechanisms.[4] When integrated early and substantively with technological development, this broad range of humanistic and social scientific perspectives do not impede research; instead, they reconfigure its trajectory, frequently generating novel research questions, strengthening institutional uptake, and enhancing long-term societal trust, thus elevating its overall quality.

Nevertheless, current modes of organizing research tend to undermine these prospects. As research for breakthrough technologies is increasingly organized through large-scale consortia and automated processes, the human dimension of inquiry, the reflective conversations, ethical hesitations, and creative digressions that once accompanied scientific revolutions in physics or genetics during the twentieth century, risk being displaced. What could once be negotiated among individual researchers now unfolds within complex systems that few can meaningfully redirect. In this sense, the scale and automation of today's technoscience make the integration of humanistic and social reflection not only desirable but urgent. Without deliberate structures that restore space for judgment, imagination, and plural perspectives,[5] we risk building technologies whose trajectory no longer reflects the societies they are meant to serve.

---

[1] Barthe, Y., Elam, M., & Sundqvist, G. (2020). Technological fix or divisible object of collective concern? Histories of conflict over the geological disposal of nuclear waste in Sweden and France. *Science as Culture* 29(2), 196–218.

[2] Wright, L., Ross, K. & Daar, A. (2005). The roles of a bioethicist on an organ transplantation service. *American Journal of Transplantation* 5(4), 821–826.

[3] Crawford, K. & Paglen, T. (2021). Excavating AI: the politics of images in machine learning training sets. *AI & Society* 36, 1105–1116.

[4] In the social sciences, the distinction between interpretative and explanatory approaches is mapped onto the twin traditions of *Verstehen* (understanding) and *Erklären* (explaining). The humanities are generally aligned within the former approach, while the natural sciences are in alignment with the latter approach. See Feest, U. (2010). *Historical Perspectives on Erklären und Verstehen*. Dordrecht: Springer.

Despite this recognition, integration of technological research and the humanities and social sciences remains uneven and frequently symbolic. Ethical, legal, and social aspects are often addressed late or not at all in the technological research process, foresight is reduced to speculative add-ons, and communication is treated as dissemination rather than as a constitutive part of knowledge production. Furthermore, graduate training still tends to reproduce disciplinary silos, while institutional incentives, merit evaluation systems and funding structures rarely reward the time, negotiation, and reflexivity required for meaningful interdisciplinary collaboration. As a result, opportunities are missed to design technologies that are not only technically sophisticated, high-performing, and scalable, but also robust, legitimate, and responsive to real-world complexity.

This perspective article argues that true technological excellence cannot be achieved without substantive integration of human and social context—not as an afterthought, but as a core criterion of research quality. We propose that excellence in breakthrough technologies should be understood as the capacity to combine technical rigor with social intelligibility, ethical robustness, and long-term relevance. Such an understanding shifts attention from isolated breakthroughs to research processes that are anticipatory, reflexive, and embedded in societal and cultural realities. It also reframes the role of humanities and social sciences from commentators or critics to co-creators of knowledge and direction.

Here we develop a structured framework across five interconnected dimensions that redefine technological excellence in contemporary research practice, as schematically illustrated in Figure 1. First, we examine ethical, legal, and social aspects not as compliance requirements, but as sources of insight that shape problem formulation and responsibility. Second, we explore the role of foresight and future imaginaries in guiding technological trajectories and aligning innovation with collective aspirations. Third, we consider graduate education as a critical leverage point for cultivating cross-disciplinary literacy and collaborative capacity. Fourth, we address visualization and communication as epistemic practices that influence how science is understood, trusted, and acted upon, both within academic institutions and in the broader communication to promote public engagement. Finally, we revisit the relationship between basic and applied research, arguing for a more integrated view of technoscience that acknowledges iterative feedback between discovery, application, and societal impact.

Across these dimensions, we address three core questions: why these practices matter for technological excellence; what limitations and barriers characterize current approaches; and how they might be redesigned to better integrate human and social dimensions into research practice. In doing so, we seek to offer concrete suggestions for how the humanities and social sciences can be more effectively integrated into the technological research process, drawing on experiences that are broadly applicable across institutional and national contexts. Our aim is not to prescribe a single model, but to articulate principles and practices that can guide the design of research environments capable of producing technologies that are groundbreaking as well as meaningful. As technological research is mobilized to confront complex global challenges, a broader conception of excellence—one that also embeds human and social context—is no longer optional but essential.

---

[5] Palmås, K. (2024). Engineering judgment and education: An Arendtian account. *Engineering Studies* 16(3), 184–205.

# 1. Ethical, legal, and social aspects

## 1.1. Ethical, legal, and social perspectives increasingly inform technological development

The relationship between technological development and the humanities and social sciences has a long history, but it has often been framed in reactive or auxiliary terms. Traditionally, ethical, legal, and social aspects (ELSA) entered the research process after technological trajectories had already been defined—typically through risk assessment, regulatory compliance, or public communication.[6,7] This model reflects a post-war understanding of scientific progress where technology was assumed to be value-neutral and societal concerns were treated as external constraints rather than internal drivers.

From the late twentieth century onward, this view has been increasingly challenged. Fields such as bioethics and science and technology studies (STS) as well as policy frameworks such as ELSA emerged in response to controversies surrounding nuclear power, biotechnology, and medical research. These approaches highlighted that technologies are never socially neutral: they embody assumptions about users, values, responsibility, and desirable futures. Ethical reflection, therefore, could not be reduced to compliance or downstream "impact".[8] More recently, this insight has been reinforced by developments in digital technologies, artificial intelligence (AI), synthetic biology, and data-intensive research, where design decisions directly shape social relations, power asymmetries, and forms of agency. In the case of generative AI, questions of alignment, guardrails, and acceptable behavior require explicit normative frameworks, highlighting how ethical reasoning, cultural interpretation, and political theory become part of the technological design process itself. In parallel, policy frameworks, particularly at the European level, have increasingly emphasized responsible research and innovation (RRI), societal desirability, and anticipatory governance.[9]

Through a parallel development, quantitative and computational social scientific approaches have developed new tools for studying the societal dimensions of technology. Econometric models may estimate the impact of industrial robots on employment and wages across local

---

labor markets.[10] Task-based frameworks can predict which occupations are most susceptible to automation.[11] Survey-based frameworks can identify the specific moral qualities—such as perceived risk, benefit, dishonesty, unnaturalness, and reduced accountability—that drive public acceptability of new technology applications. Mathematical models of norm change can predict shifts in public moral opinion on technology-related issues across dozens of societies. Historical analysis shows that the distributional effects of technology depend critically on institutional and regulatory choices, not on technological properties alone.[12]

The state of the art thus recognizes that humanities and social sciences are not merely observers of technological change but are essential to understanding how technologies come to matter in the world.

## 1.2. Current integration practices fall short of their transformative potential

Currently, the integration of humanities and social sciences into cutting-edge technological research is widely acknowledged as necessary yet remains uneven and fragile. On the opportunity side, there is growing awareness among funders, institutions, and researchers that ethical, legal, and social dimensions affect research excellence itself, not just legitimacy or acceptance. Technologies that ignore societal contexts risk failure, public backlash, or harmful consequences. Conversely, early engagement with normative questions can open new research directions, refine problem definitions, and strengthen long-term robustness. This is particularly salient for groundbreaking technologies, where uncertainty, scale, and irreversibility are high.

Despite widespread recognition of the value of integrating the humanities and social sciences into technological research, integration often remains superficial. Humanities and social sciences scholars are commonly brought in late and expected to manage ethical implications or enhance societal legitimacy, rather than shape research priorities from the outset. This positioning reduces their role to reactive problem-solving instead of constitutive participation. Differences in epistemic cultures further complicate collaboration, as contrasting standards of evidence, temporal horizons, and criteria for success can marginalize qualitative, interpretive, or normative forms of inquiry. Expectations may also diverge, with technical researchers seeking clear recommendations or solutions, while humanities and social science perspectives emphasize ambiguity, critical reflection, and the contestability of values. These tensions are reinforced by structural conditions, including funding mechanisms, evaluation practices, and project timelines that prioritize rapid technical deliverables over slower, deliberative forms of knowledge production.

The result is a paradoxical situation: while the need for integration is rhetorically accepted, actual collaboration is often shallow, symbolic, or confined to peripheral work packages. The history of genetically modified crops offers a cautionary example.[13] In several regions, public engagement and social analysis were introduced only after technological pathways and commercialization strategies had already been defined. As a result, debates became polarized around risk and trust rather than shaped collaboratively around agricultural priorities, food systems, and distributional effects. Earlier and more substantive integration of social and ethical inquiry might have reframed both research agendas and public reception.[14]

---

[10] Acemoglu, D., & Restrepo, P. (2020). Robots and jobs: Evidence from US labor markets. *Journal of Political Economy* 128(6), 2188–2244.

[11] Autor, D. H. (2015). Why are there still so many jobs? The history and future of workplace automation. *Journal of Economic Perspectives* 29(3), 3–30.

[12] Acemoglu, D. (2024). *Institutions, Technology, and Prosperity.* Nobel Prize lecture. Nobel Prize Outreach.

## 1.3. Embedding humanities and social sciences will enable co-production of technological knowledge

Moving from diagnosis to action requires shifting attention from abstract arguments about the value of interdisciplinarity to the concrete conditions under which it can be made to work in practice. While the case for integrating humanities and social sciences into the development of groundbreaking technologies is increasingly well articulated, far less consensus exists on how such integration should be organized, sustained, and evaluated across different institutional and national contexts. Persistent challenges—including disciplinary asymmetries, misaligned incentives, and the tendency to treat societal perspectives as auxiliary rather than constitutive—underscore the need for practical, scalable approaches that embed humanistic and social inquiry directly into research processes. The following recommendations therefore focus on actionable strategies that can be adopted across research systems to support genuine co-production of knowledge and to enhance both scientific excellence and long-term societal relevance:

- **Integrate the humanities and social sciences already at the stage of agenda setting and project design**, for example, by involving relevant scholars in the formulation of research questions, challenge definitions, and success criteria before proposals are finalized. Early engagement ensures that normative, historical, and societal considerations contribute to shaping the direction of research rather than merely being later add-ons. This can be realized through co-designed calls, interdisciplinary scoping workshops, or cross-disciplinary concept notes at the pre-proposal stage.

- **Assign humanities and social science researchers formal roles within core research teams**—such as co–principal investigators, work-package leaders, or long-term embedded collaborators—rather than limiting their participation to advisory boards or short-term consultations. This helps ensure that societal, ethical, and historical perspectives shape ongoing technical decisions rather than post hoc assessments. Such structural inclusion indicates that interdisciplinary collaboration is a core dimension of research quality, not an afterthought.

- **Create shared research practices and spaces**, including joint seminars, lab rotations, co-supervised doctoral projects, and collaborative writing efforts, where technical and non-technical researchers engage with the same empirical material, data, or prototypes and develop a common scientific language. Such practices foster mutual literacy and reduce disciplinary compartmentalization. Notably, this requires long-term engagement from the humanities and social science scholars, enabling them to forge personal and professional networks, as well as specific field knowledge regarding particular technologies. Without this continuity, interdisciplinary collaboration risks remaining superficial or episodic.

---

[13] Levidow, L. & Carr, S. (2007). GM crops on trial: technological development as a real-world experiment. *Futures* 39(4), 408–431.

[14] Wilsdon, J. & R. Willis (2004). *See-Through Science: Why Public Engagement Needs to Move Upstream.* London: Demos.

- **Support sustained, long-term engagement throughout the full research lifecycle**, from early exploratory phases to prototyping, deployment, and evaluation. This can include regular cross-disciplinary reflection checkpoints, iterative impact assessments, and adaptive governance structures that evolve as technologies and contexts change. Incorporating such continuity ensures that ethical, legal, and social considerations remain integral to research development rather than being confined to isolated stages or final reporting requirements.

- **Align evaluation, promotion, and funding criteria with interdisciplinary contributions** by explicitly recognizing co-authored outputs, process-oriented research, conceptual advances, and societal relevance alongside technical performance metrics. Without institutional incentives that acknowledge and value such work, meaningful interdisciplinary integration remains fragile and is unlikely to persist beyond individual projects.

- **Treat normative disagreement and epistemic tension as productive research inputs**, and not as obstacles to be minimized. Structured formats such as deliberative workshops, documented dissent in project reports, or reflexive impact statements can help teams surface and work through conflicting assumptions about risk, value, responsibility, and desired futures.

- **Balance between distributed and centralized modes of organization.** On the one hand, there is a strong case for embedding ethical, legal and social science scholars within interdisciplinary teams that address key challenges regarding specific groundbreaking technologies. On the other hand, there is equally a need for the founding of central nodes, such as institutional hubs or competence centers, which refer natural scientists to ELSA-related researchers to consolidate expertise, provide methodological guidance, ensure quality standards, and serve as advisory bodies to policymakers and funding agencies.

## 2. Foresight and future imaginaries in research excellence

### 2.1. Technological development is guided by implicit futures

Technological research is never oriented solely toward the present. It is animated by implicit or explicit imaginaries of the future: visions of desirable societies, anticipated risks, emerging markets, geopolitical positioning, or transformative breakthroughs.[15] These imaginaries influence what problems are considered urgent, which research pathways appear promising, and what forms of innovation are deemed legitimate. Yet, such future orientations often remain implicit, embedded in funding calls, roadmaps, mission statements, and narratives of competitiveness or sustainability.

---

[15] Jasanoff, S., & Kim, S.-H. (2015). *Dreamscapes of Modernity: Sociotechnical Imaginaries and the Fabrication of Power*. Chicago, IL.: University of Chicago Press.

Historically, large-scale scientific projects have been shaped by powerful collective imaginaries of modernization, national security, economic growth, or human progress.[16] In many cases, these visions were articulated primarily within technical or political domains, with limited reflexive examination of their underlying assumptions.[17] Today, as research increasingly addresses complex global challenges—such as climate change, demographic shifts, digital transformation, and planetary health—the stakes of these future orientations are higher and more contested. Competing visions of the future coexist, and technological trajectories can reinforce particular social, economic, or geopolitical configurations.

Again, the humanities and social sciences represent a wide palette of approaches to engage with futures. Generally, foresight practices are either based on making predictions or projections using existing data, or on exercises such as scenario-planning which expands the range of considered futures. These, in turn, map onto the distinction between the above-mentioned explanation and understanding traditions. Within explanation-oriented social science, quantitative and computational social scientific approaches have developed increasingly sophisticated foresight capabilities, offering falsifiable predictions about the labor market effects of automation,[18] the distributional consequences of technological change for wages and inequality,[19] the moral acceptability of new technological applications,[20] how public opinion on technology-related moral issues will evolve,[21] and which adoption pathways are most likely under different conditions.[22]

Moreover, within the understanding-based interpretive traditions of humanities and social sciences, there is also considerable interest in futures. Here, the question is less about making predictions, but rather to offer analytical and creative tools to examine and shape future imaginaries. Fields such as science and technology studies, political theory, anthropology, history, design futures, futures studies, and even science fiction have long demonstrated that expectations about the future can become self-fulfilling: they mobilize resources, attract investment, structure regulation, and shape institutional priorities.[23] From these perspectives, foresight is not a neutral forecasting exercise, but a normative and political practice that shapes technological pathways.[24]

---

[16] Latour, B. (1987). *Science in Action: How to Follow Scientists and Engineers Through Society*. Cambridge, MA.: Harvard University Press.

[17] Mazé, R. (2019). Politics of designing visions of the future. *Journal of Futures Studies* 23(3), 23–38.

[18] Frey, C. B., & Osborne, M. A. (2017). The future of employment: How susceptible are jobs to computerisation? *Technological Forecasting and Social Change* 114, 254–280.

[19] Acemoglu, D., & Johnson, S. (2023). *Power and Progress: Our Thousand-Year Struggle Over Technology and Prosperity*. PublicAffairs.

[20] Eriksson, K., Strimling, P., & Vartanova, I. (2025). What makes AI applications acceptable or unacceptable? A predictive moral framework. arXiv 2508.19317.

[21] Strimling, P., Vartanova, I., & Eriksson, K. (2022). Predicting how US public opinion on moral issues will change from 2018 to 2020 and beyond. *Royal Society Open Science* 9(4), 211068.

[22] Guo, Y., Yousef, E., & Naseer, M. M. (2025). Examining the drivers and economic and social impacts of cryptocurrency adoption. *FinTech* 4(1), 5.

[23] van der Helm, R. (2009). The vision phenomenon: Towards a theoretical underpinning of visions of the future and the process of envisioning. *Futures* 41(2), 96–104.

[24] Robinson, J. (1988). Unlearning and backcasting. *Technological Forecasting and Social Change* 39(5), 325–338.

The space race of the mid-twentieth century illustrates this self-fullfilling dimension.[25] Visions of technological supremacy and national prestige mobilized vast public investment, reoriented educational systems, and accelerated developments in computing and materials sciences. These imaginaries did not merely describe a future; they actively organized institutions and priorities in the present towards such ends as a form of prefiguration.[26] Contemporary narratives surrounding artificial intelligence, synthetic biology,[27,28] or quantum technologies[29] function in similar ways, shaping funding landscapes and geopolitical competition before specific applications are fully defined. In particular, narratives about transformative artificial intelligence—from utopian visions of automated abundance to fears of systemic disruption—play a powerful role in shaping research agendas, investment flows, and regulatory debates. Understanding and critically examining these technological imaginaries is therefore essential for steering innovation toward socially desirable outcomes. Recognizing this aspect repositions foresight from a peripheral planning tool to a constitutive dimension of research excellence. If future imaginaries shape what is funded, legitimized, and institutionalized, then making them explicit and subject to critical scrutiny becomes a core research responsibility. Excellence, in this sense, cannot be defined solely by technical sophistication or short-term performance metrics; it must also encompass reflexive engagement with the futures being enacted through research practice. Integrating structured foresight and critical reflection into technological research allows competing visions to be examined, underlying assumptions to be surfaced, and alternative trajectories to be explored. In doing so, research systems move from passively inheriting dominant narratives to actively shaping more robust, inclusive, and socially responsive technological futures.

## 2.2. Current foresight practices are often narrow, instrumental, and short-term

Despite the growing prominence of foresight language in research policy, current practices frequently remain limited in scope. In many settings, foresight is treated as a strategic forecasting exercise focused on market opportunities, geopolitical positioning, or competitive advantage under a logic of risk-reduction, planning, and control. Roadmapping and horizon scanning are often conducted within disciplinary or sectoral silos, emphasizing technological feasibility and economic scalability while giving less attention to broader societal transformations.[30] Addressing this gap requires greater attention to open-ended complexity, radical uncertainty, politics, and power. Furthermore, a too-narrow approach to forecasting tends to reduce the future into a space filled with present-day risks and economic interests, instead of considering the more long-term effects of current actions.[31] This narrow framing risks conflating technological acceleration with societal progress. When future imaginaries are defined primarily in terms of growth, speed, or disruption, alternative futures centered on resilience, equity, justice, care, or sustainability may be marginalized.[32] Moreover, foresight exercises often privilege short- to medium-term horizons aligned with funding cycles or political mandates, limiting consideration of long-term systemic consequences.[33]

---

[25] Davis Cross, M.K. (2019). The social construction of the space race: then and now, *International Affairs* 95(6), 1403–1421.

[26] Sareen, S., & Juhola, S. (Eds.). (2026). *Societal Transitions to Sustainability: The Prefigurative Politics of Present Transformation*. Springer Nature Switzerland.

[27] Groff-Vindman, C.S., Trump, B.D., Cummings, C.L. et al. (2025). The convergence of AI and synthetic biology: the looming deluge. *npj Biomedical Innovations* 2(20)

[28] Hynek, N. (2026). Synthetic biology/AI convergence (SynBioAI): security threats in frontier science and regulatory challenges. *AI & Society* 41, 951–968.

[29] Suter, V., Pöhlmann, G., Ma, C. *et al.* (2026). Quantum technologies and geopolitics: comparing parliamentary rhetoric. *EPJ Quantum Technologies* 13(10).

[30] Avelino, F., Wijsman, K., Van Steenbergen, F., Jhagroe, S., Wittmayer, J., Akerboom, S., Bogner, K., Jansen, E. F., Frantzeskaki, N., & Kalfagianni, A. (2024). Just sustainability transitions: politics, power, and prefiguration in transformative change toward justice and sustainability. *Annual Review of Environment and Resources* 49(1), 519–547.

Another limitation lies in participation. Foresight processes are frequently confined to experts within specific domains, with limited involvement of diverse publics or interdisciplinary perspectives.[34] This can reinforce epistemic hierarchies and reduce the capacity to anticipate unintended consequences, value conflicts, biases, or distributional effects. In areas such as artificial intelligence, biotechnology, and climate technologies, early design choices can lock in infrastructures and norms that are difficult to reverse. Without broader and more reflexive foresight practices, research excellence risks becoming technically sophisticated but socially brittle.

These limitations stem less from lack of awareness than from the institutional incentives that shape research practice. Funding structures, evaluation criteria, and organizational cultures often reward rapid output, novelty, and competitive positioning over long-term reflection and pluralistic engagement. As a result, foresight becomes an add-on, possibly obligatory section in grant proposals, rather than a structured, deliberative practice integrated into research design.

## 2.3. Embedding plural and reflexive foresight in research design

Moving from recognition to realization requires institutionalizing foresight as a core dimension of research design rather than treating it as a predictive or strategic add-on. If technological excellence is to be socially robust and future-oriented, foresight must become a structured, interdisciplinary, and reflexive practice embedded throughout the research lifecycle. This entails shifting from narrow forecasting toward plural and deliberative approaches that explicitly examine how technological trajectories are imagined, justified, and governed. The following strategies outline practical ways to embed such foresight across research systems:

- **Integrate structured future-oriented reflection at the agenda-setting stage.** Research programs should not only ask what is technically feasible or economically competitive, but also which futures their work implicitly advances and which alternatives it sidelines, including critical reflection on what types of institutional changes or conditions that are necessary to move from desirability to viability.[35] Interdisciplinary scenario workshops, narrative prototyping, and anticipatory exercises can help teams articulate and challenge underlying assumptions before research directions become locked in.[36,37]

---

[31] Adam, B., & Groves, C. (2007). *Future matters: Action, knowledge, ethics* (Vol. 3). Brill.

[32] Fridolfsson, C., Gislén, Y., Mukhtar-Landgren, D., Pettersson, C., Rahm, L., & Ståhl, Å. (2025). Välkommen till framtiden. *Fronesis* 86-87, 6–25.

[33] Stirling, A. (2024). Responsibility and the hidden politics of directionality: Opening up 'innovation democracies' for sustainability transformations. *Journal of Responsible Innovation* 11(1), 2370082.

[34] Lundberg, R., Pink, S., & Pinyon, Z. (2025). Interdisciplinary futures? A conceptual approach. *Futures* 172, 103648.

[35] Wright, E.O. (2010). *Envisioning Real Utopias.* London: Verso.

- **Embed humanities and social science scholars as co-designers of foresight processes**, rather than positioning them as external commentators. Institutions should involve humanities and social sciences in defining research challenges, evaluating long-term implications, and interrogating the metaphors and narratives that shape technological visions. Such collaboration enables examination of historical precedents, political imaginaries, and value commitments that would otherwise remain implicit. At the same time, meaningful co-design requires that humanities and social sciences scholars develop substantive familiarity with the relevant technologies through sustained engagement, shared empirical work, and technical literacy-building. Reciprocal competence—normative and historical insight on the one hand, technological understanding on the other—is essential for foresight to function as an integrative rather than parallel activity.

- **Complement interpretive, imagination-based foresight approaches with data-driven predictive approaches** that model technology diffusion, assess public acceptability of technology through surveys, and predict norm change. For example, frameworks that identify the specific moral qualities driving public resistance to AI applications can be used prospectively to anticipate which emerging technologies are likely to face acceptance challenges. Moreover, historical analysis of how institutional choices shaped the distributional outcomes of past technologies can inform governance frameworks for emerging ones. By integrating these approaches, both two dominant traditions in foresight are utilized.

- **Explicitly address distributional and geopolitical dimensions in foresight practices.** Evaluations of emerging technologies should systematically consider who benefits, who bears risks, and how global inequalities may be reinforced or mitigated. Incorporating perspectives from political economy, development studies, and postcolonial scholarship can broaden the scope of anticipated impacts beyond immediate markets or national interests.

- **Protect long-term horizons institutionally** by using dedicated and sustainable funding mechanisms, periodic anticipatory reviews, and evaluation criteria that recognize long-term societal relevance and can counterbalance short-term performance pressures. Large research consortia and excellence clusters are particularly well positioned to sustain such reflection capitalizing on their scale and internal diversity of expertise.

- **Recognize normative disagreement as a productive resource.** Competing future imaginaries should not be treated as obstacles to coherence, but as opportunities for critical learning.[38] Structured deliberation formats—such as documented dissent, parallel scenario development, or reflexive impact statements—can surface conflicting assumptions and strengthen the robustness of research strategies.

---

[36] Light, A. (2021). Collaborative speculation: Anticipation, inclusion and designing counterfactual futures for appropriation. *Futures* 134, 102855.
[37] Vervoort, J. M., Bendor, R., Kelliher, A., Strik, O., & Helfgott, A. E. R. (2015). Scenarios and the art of worldmaking. *Futures* 74, 62–70.
[38] Engeström, Y., & Sannino, A. (2021). From mediated actions to heterogenous coalitions: Four generations of activity-theoretical studies of work and learning. *Mind, Culture, and Activity* 28(1), 4–23.

- **Ensure participatory and cross-sectoral engagement in defining technological futures** by including, where appropriate, diverse societal actors, such as civil society organizations, professional communities, and affected groups. Early engagement can expand the range of considered futures and enhance legitimacy and relevance without subordinating scientific autonomy to immediate public opinion.

# 3. Graduate education as a leverage point for integrating humanities and social sciences

## 3.1. Leveraging future research cultures through doctoral education

Graduate education is a primary site for shaping of future research cultures. Indeed, doctoral programs are not merely sites of knowledge transmission; they are formative environments where epistemic norms, methodological standards, professional identities, and collaborative habits are cultivated. The structures and incentives embedded in graduate training shape how future researchers understand excellence, define problems, and engage across disciplinary boundaries. Therefore, graduate education may be used as a leverage point for integrating the humanities and social sciences in technological research excellence—one that sits alongside policy, funding and research infrastructures. As a leverage point for cross-disciplinary exchange, doctoral training may also be understood in the context of broader policy goals of life-long learning.[39]

Traditionally, doctoral education has usually been organized along disciplinary lines. This model has proven highly effective for cultivating depth of expertise and methodological rigor. However, as research increasingly addresses complex societal challenges and technologically mediated transformations, disciplinary specialization alone may be insufficient. Breakthrough technologies often require collaboration across domains with different epistemic cultures, temporalities, and evaluative criteria. The ability to navigate such differences is not an automatic by-product of disciplinary excellence; it is a skill that must be learned.

Recent developments in artificial intelligence further reinforce this need for cross-disciplinary literacy. Interaction with increasingly capable AI systems—such as large language models used in research, analysis, and programming—often takes place through natural language prompts and interpretive dialogue. Competences cultivated in the humanities, including linguistic precision, rhetorical framing, and sensitivity to ambiguity, therefore become increasingly relevant within technological workflows themselves. Graduate education that integrates these interpretive skills alongside technical training can better prepare researchers for emerging AI-mediated research environments.

The humanities and social sciences contribute distinctive forms of literacy to this process: conceptual analysis, historical contextualization, normative reasoning, interpretive sensitivity, and reflexive critique. Conversely, technological fields offer modes of experimentation, prototyping, and quantitative modeling that can enrich inquiry in the humanities and social sciences. Graduate education represents a uniquely dynamic stage where such cross-disciplinary literacies can be cultivated before professional identities and institutional trajectories become more fixed. Integrating these perspectives at the doctoral level should therefore be a structural intervention in the future organization of research. Here, there are ongoing initiatives to learn from: One example is the Swedish National Doctoral School in Digital Humanities (DASH), which provides structured, cohort-based training at the intersection of ICT (Information and Communication Technology), culture, and society.[40]

---

[39] Foka, A., Gulliksen, J., Heinz, F., & Loufti, A. (2025). Advanced Digital Skills 2035: Future Scenarios and Implications for Policy Making on Higher Education for Digital Skills and Lifelong Learning. *EDULEARN25 Proceedings*, 5032–5040.

Historical precedents suggest that such integration can reshape entire research cultures. For example, the post-war reorganization of biomedical research in several countries linked laboratory science more closely with clinical training and public health systems, producing new institutional forms such as translational research hospitals and interdisciplinary medical faculties.[41] These reforms did not dilute disciplinary depth; rather, they created environments in which conceptual and practical questions evolved together.

## 3.2. Structural barriers limit meaningful interdisciplinary training

Despite growing recognition of the need for interdisciplinary competence, graduate education remains largely siloed. Arguably, current doctoral structures systematically disincentivize meaningful interdisciplinarity. Doctoral programs are typically governed by disciplinary curricula, evaluation standards, and supervisory traditions that reward depth within established fields. While interdisciplinary courses and workshops are increasingly offered, they are often peripheral to the core requirements of doctoral training and may lack continuity.

Short-term exposure, such as isolated lectures or elective modules, rarely suffices to foster durable interdisciplinary competence. Differences in epistemic culture can create additional challenges: standards of evidence, authorship practices, publication formats, and time horizons vary significantly across fields. Without sustained engagement, these differences can reinforce misunderstandings or tokenistic collaboration rather than genuine integration.

Administrative structures may also constrain innovation: joint supervision across faculties can be complex, credit allocation systems may not recognize interdisciplinary coursework, and evaluation criteria for doctoral theses often remain discipline specific. In such contexts, doctoral candidates may perceive interdisciplinary engagement as risky, particularly if it complicates timely completion or weakens alignment with established career paths. This problem also extends beyond the PhD to evaluation and hiring criteria, for instance in relation to postdoctoral selection, promotion, and tenure.

These structural conditions generate what may be described as a "graduate-school paradox". Doctoral education is widely recognized as a promising site for integrating technological research with the humanities and social sciences, yet institutional arrangements often confine such efforts to symbolic or peripheral initiatives. Although doctoral training is structurally central to the reproduction of research cultures, it remains operationally marginal to meaningful interdisciplinary integration. At the same time, many natural sciences and technology disciplines face a growing challenge in maintaining sufficient intradisciplinary depth. As fields become more complex and technically demanding, doctoral programs have increasingly shifted toward structured coursework and compressed timelines—particularly within three-year PhD formats—leaving less space for exploratory, individually-driven research. Under these conditions, interdisciplinary components may be perceived as competing with, rather than complementing, disciplinary rigor. If doctoral education is to function as a genuine leverage point for reshaping research cultures, integration must therefore be deliberately designed to enrich, rather than dilute, disciplinary depth. The aim is not to replace specialized expertise with generalist breadth, but to cultivate technically competent researchers who are also capable of reflexive, interdisciplinary engagement.

---

[40] La Mela, M., Brodén, D., Cocq, C., Foka, A., Golub, K., LaMonica, C., & Westin, J. (2024). DASH Swedish National Doctoral School in Digital Humanities : From Local Expertise to National Research Infrastructure. Proceedings of the Huminfra Conference (HiC 2024), 110–114.

[41] Crabu, S. (2018). Rethinking biomedicine in the age of translational research: organisational, professional, and epistemic encounters. *Sociology Compass* 12, e12623.

## 3.3. Designing graduate education for cross-disciplinary literacy and collaboration

Strengthening the role of graduate education in integrating humanities, social sciences, and technological research requires more than curricular reform. Cross-disciplinary literacy does not emerge from occasional exposure or short-term workshops; it develops through sustained physical and social embedding in multiple research environments. Doctoral researchers must experience different disciplines operating in practice: how problems are framed, how data are interpreted, how disagreement is handled, and how success is evaluated. Such embedding allows candidates to acquire not only conceptual understanding but also tacit knowledge of distinct epistemic cultures. The following strategies outline how institutions can implement this form of integration concretely:

- **Recognize social integration as part of research training.** Informal interaction—shared seminars, interdisciplinary reading groups, social events within clusters, joint coffee and lunch breaks—plays a critical role in building intellectual trust. Institutions should value and facilitate these arenas, as they often enable the kinds of candid conversations in which new questions and collaborations emerge.

- **Require structured periods of physical co-location across research environments.** Doctoral candidates should spend extended time—weeks or months rather than days—embedded in partner departments, laboratories, archives, or design studios outside their home discipline. Formal rotation schemes, visiting doctoral fellowships within excellence clusters, and dual institutional affiliations can ensure that such exchanges are systematic rather than ad hoc.

- **Establish joint workspaces and shared infrastructures.** Co-location is most effective when supported by shared physical spaces—common coffee/lunch rooms, offices, project rooms, or collaborative labs—where doctoral researchers from different disciplines meet and work side by side. Spatial proximity fosters informal exchange, lowers barriers to consultation, and enables the gradual development of mutual understanding.

- **Implement sustained co-supervision anchored in everyday practice.** Joint supervision should not be limited to periodic meetings. Supervisors from different disciplines should participate actively in each other's research environments—attending lab meetings, seminars, or reading groups—to ensure that doctoral candidates are not navigating incompatible expectations in isolation. Clear agreements about evaluative standards and thesis formats are essential to avoid placing students between conflicting criteria.

- **Create long-term interdisciplinary cohorts rather than isolated collaborations.** Graduate schools or thematic doctoral programs should organize candidates into stable cohorts that meet regularly over multiple years. Shared seminars, annual retreats, and collaborative research design workshops can build trust and common vocabulary over time, reducing the risk of superficial engagement.

- **Integrate collaborative research outputs into doctoral milestones.** Requirements such as joint conference presentations, collaborative project proposals, or co-authored publications, would anchor interdisciplinary work within formal progression structures. This signals that cross-disciplinary engagement is integral to doctoral development rather than an extracurricular activity. At the same time, while it may be neither feasible nor appropriate to require co-authored publications or joint proposals as a condition for passing individual thesis evaluations, doctoral schools can nonetheless raise expectations for how candidates demonstrate interdisciplinary insight in their thesis synthesis by articulating a deeper understanding of the interplay between humanities and social sciences and technological development, demonstrating reflexive awareness, conceptual integration, and methodological implications.

- **Align funding structures with sustained embedding.** Stipends, travel grants, and workload allocations must explicitly support extended exchanges across environments. Without dedicated resources, physical embedding remains aspirational and unevenly distributed.

## 4. Communication and visualization as epistemic and societal practices

### 4.1. Visualization shapes what scientists and societies are able to see

Scientific communication is often described as a matter of simplification: the translation of complex findings into accessible formats for broader audiences. As such, communication is seen to be unidirectional, reporting on events from the technosciences. While this function is essential, it captures only part of its significance. Communication is not merely a downstream activity; it is constitutive of scientific knowledge and of its societal meaning.

The same holds true for the particular ways in which research data are rendered visible. Visualizations generated through science shape how scientists interpret phenomena. They also shape how societies understand, trust, contest, and act upon scientific claims.[42]

---

[42] Foka, A., & Von Bonsdorff, J. (2025). *AI and Image: Critical Perspectives on the Application of Technology on Art and Cultural Heritage.* Cambridge: Cambridge University Press.

Imaging as data production, measurement, and documentation is intrinsically entwined with cultural ideas about epistemology, truth, and veracity.[43] Indeed, a visualization is not merely a passive mirror that neutrally represents the world, but an event[44]—or an act—of imagination, documentation and communication.

Throughout the history of science, new visualization techniques have opened access to previously inaccessible aspects of reality, transforming both intradisciplinary research practice and public imagination. The double helix model of DNA did not simply illustrate an already understood molecular structure; it helped stabilize and disseminate a particular interpretation of biological organization, reshaping conceptions of heredity, identity, and agency. The accompanying metaphor of a "code" further reinforced a view of heredity as programmable and deterministic, influencing both scientific reasoning and broader cultural narratives about identity and control.[45]

Scientific photography in twentieth-century physics made invisible or very fast phenomena perceptible and reproducible, altering standards of evidence.[46] More recently, the first image of a black hole was not merely a technical milestone; it became a global cultural event, shaping how fundamental physics is imagined beyond specialist communities.[47] Similarly, digital 3D imaging in archaeology has reconfigured how objects, bodies, and sites are interpreted[48]—within expert communities and in courtrooms, museums, and public debates.

In each case, visualization techniques did not simply display knowledge; they structured the epistemic field within which knowledge was produced and circulated. Choices about scale, color, dimensionality, and framing influence what patterns are perceived and which interpretations appear plausible. As new instruments generate increasingly complex, high-dimensional datasets—from correlated time- and energy-resolved measurements to advanced microscopy, space observatories and machine-learning-driven simulations—the challenge intensifies. In this way, visualization serves as a critical interface between measurement and meaning.

At the same time, visualizations shape societal expectations of science. Images can stabilize trust or generate controversy. They can simplify complexity responsibly or obscure uncertainty. They can empower publics to engage with evidence or create false impressions of certainty. In an era of AI-assisted reconstruction and immersive digital environments, the distinction between observation, simulation, and interpretation becomes increasingly blurred. Visualization is therefore both an epistemic and a civic practice.

Visual language is inseparable from written and oral language. The words, metaphors, and narrative frames used to describe scientific findings shape perception as profoundly as images do.[49] Concepts such as "genetic code", "editing", "intelligence", "learning", or "groundbreaking technologies" carry implicit analogies that guide interpretation. They can suggest determinism, agency, inevitability, or disruption, even when the underlying phenomena are more contingent or relational. Just as images foreground certain features and background others, terminology directs attention and frames expectations—note for instance the case of wave–particle duality in quantum mechanics. Just like language, visualization is therefore not a neutral vehicle for communication but an integral component of how scientific realities are constructed and understood.

---

[43] Daston, L., & Galison, P. (2007). *Objectivity*. New York, NY: Zone Books.

[44] Azoulay, A. (2015). *Civil Imagination: A Political Ontology of Photography*. London: Verso.

[45] Fox Keller, E. (1995). *Refiguring Life: Metaphors of Twentieth-Century Biology*. New York, NY.: Columbia University Press.

[46] Penny, D. (2025). Thoughtful photography for scientists. *Nature Reviews Physics* 7, 466–467.

[47] Sunde, E.K. (2024). From outer space to latent space. *Philosophy of Photography* 15(1-2: Expanded Visualities: Photography and Emerging Technologies), 123–142.

[48] Pendić, J. & Molloy, B. (2024). The Use of 3D Documentation for Investigating Archaeological Artefacts. In: Hostettler, M., Buhlke, A., Drummer, C., Emmenegger, L., Reich, J. & Stäheli, C. (eds) *The 3 Dimensions of Digitalised Archaeology.* Springer, Cham.

The humanities and social sciences, alongside artistic and design disciplines, contribute essential expertise to this domain. They bring tools for analyzing representation, metaphor, narrative framing, aesthetics, and the politics of images. Integrating these perspectives strengthens both scientific interpretation and democratic engagement.

### 4.2. Visualization remains structurally undervalued

Despite its epistemic and societal importance, visualization is frequently treated as a technical service or as a late-stage dissemination tool. In many research environments, visual design is delegated to specialists after analytical decisions have already been made. This separation risks obscuring the interpretive work embedded in visual choices and underestimating their influence on scientific reasoning.

As data-intensive research expands, visualization challenges grow more complex. Integrating heterogeneous datasets, representing uncertainty, or conveying probabilistic outcomes requires interdisciplinary expertise. While there are examples of infrastructures that offer services to meet these challenges—note, for instance, the Swedish InfraVis[50]—institutional incentives rarely reward sustained attention to representational design, even when it fundamentally improves interpretability.

The societal dimension is equally underdeveloped. Public communication efforts often prioritize simplification and outreach metrics without systematically reflecting on how images frame technological futures and shape public understanding. Visualizations can reinforce narratives of inevitability, control, or disruption without acknowledging contingency and uncertainty. Terminology can have similar effects. Describing technologies as "revolutionary", "disruptive", or "intelligent" may amplify expectations, compress perceived timelines, or obscure unresolved uncertainties. Conversely, overly technical or opaque language may distance the public from meaningful engagement. The framing of scientific developments through words and images thus shapes not only understanding, but also trust, political response, and investment priorities. In politically sensitive domains—such as climate modeling, genetic engineering, clinical imaging, or artificial intelligence—visual representations can shape regulatory debates and public trust by framing what counts as evidence, risk, and responsibility.

---

[49] Currie, J.S.G. & Clarke, B. (2022). Fighting talk: the use of the conceptual metaphor climate change is conflict in the UK Houses of Parliament, 2015–2019. *Journal of Language and Politics* 21, 589–612.
[50] Weinkauf, T., Romero, M., Kerren, A., Larsson, E., Latino, F. et al. (2025). InfraVis: The Swedish Research Infrastructure for Visualization Support. In: C. Gillmann, M. Krone, G. Reina, T. Wischgoll (ed.), *VisGap: The Gap between Visualization Research and Visualization Software.* Proceedings, The Eurographics Association.

Climate modeling offers a particularly clear example. The term "tipping point" or the iconic "hockey stick" graph, representing global temperature change, became both scientific synthesis and political symbols.[51] Its visual clarity strengthened public awareness of anthropogenic warming, but it also became a focal point of political contestation, illustrating how graphical representation can concentrate epistemic authority and political contestation simultaneously.[52] Here, there are multiple examples from biomedical research on cloning and gene editing: Dolly the sheep; fluorescent genetically modified fish; and CRISPR-edited "muscled" animals.

The rapid diffusion of generative AI tools adds further complexity.[53] Synthetic images, simulations, textual explanations, and automated visual analytics increasingly mediate how scientific results are generated, interpreted, and communicated. This blurs boundaries between measurement, reconstruction, and algorithmic artefact. Without explicit standards for transparency and documentation, both scientists and publics may struggle to assess the epistemic status of such outputs.

If excellence in groundbreaking technologies is to include societal robustness and epistemic clarity, visualization must be recognized as a structural component of research practice rather than relegated to the status of auxiliary support.

## 4.3. Embedding communication and visualization across research and society

Recognizing communication and visualization as both an epistemic and societal practice requires integrating expertise throughout the research lifecycle. These activities should be treated as processes where scientific interpretation, technological innovation, and public meaning intersect. The following strategies outline how this integration can be implemented concretely:

- **Integrate communication and visualization expertise at the earliest stages of research design.** Artists, writers, designers, data visualization specialists, and communication and visualization scholars should participate when new instruments, imaging technologies, science communication or analytical pipelines are developed. Early collaboration ensures that visual strategies support exploratory analysis while anticipating how results may be interpreted beyond specialist communities.

- **Establish interdisciplinary communication and visualization hubs with sustained engagement.** Dedicated infrastructures that combine computational, aesthetic, and critical competencies should be structurally linked to research clusters. Long-term partnerships enable visualization to shape both scientific interpretation and public communication strategies.

- **Document and communicate the construction of visualizations transparently.** Research teams should articulate how images are generated, including algorithmic choices, reconstruction techniques, and aesthetic decisions. Making these processes visible strengthens internal reflexivity and supports public trust, particularly in AI-assisted or simulation-based representations.

---

[51] Barber, G. (2025) The Math of Catastrophe. *Quanta Magazine*. September 15. https://www.quantamagazine.org/the-math-of-climate-change-tipping-points-20250915/

[52] See also *The Economist* magazine's 2019 editorial on the "Warming stripes" graphic: https://www.economist.com/leaders/2019/09/19/the-climate-issue

[53] McQuire, S., Pfefferkorn, J., Sunde, E. K., Lury, C., & Palmer, D. (2024). Seeing photographically. *Media Theory* 8(1), 1–18.

- **Develop visual literacy training for researchers and the public.** Study programs and professional development initiatives should address how visual expressions influence interpretation, how uncertainty can be represented responsibly, and how emerging technologies reshape the status of images. Outreach initiatives can extend such literacy beyond academia, fostering informed public engagement with scientific imagery.

- **Cultivate reflexive language practices alongside visual literacy.** Research teams should critically examine the metaphors, labels, and narrative frames used to describe their work, particularly in emerging technological domains. Structured reflection on terminology—for example, through interdisciplinary workshops or collaborative writing practices—can help prevent misleading analogies and foster more precise, responsible communication. One outcome of such exercises may be scholarly essays that combine literary text with illustrations and artwork.

- **Encourage collaborative and experimental visual practices.** Interdisciplinary studios, co-design workshops, sci-art collaborations,[54] and partnerships with artists and cultural institutions can foster innovative and ethical approaches to representing complex phenomena. Such formats create space for questioning established visual conventions and exploring alternative modes of communication and visualization that resonate across societal contexts.

- **Align evaluation criteria with epistemic and communicative innovation.** Institutions should recognize that breakthroughs may arise not only from new data but also from novel ways of rendering data interpretable and socially meaningful. Contributions to visualization methodology, transparency standards, and public engagement should be valued alongside technical findings.

# 5. Rethinking the relationship between basic and applied research

## 5.1. The boundary between basic and applied research is historically contingent

Science has long been labelled as either basic or applied—a categorization that (not least in the North American context) owes a lot to Vannevar Bush's landmark 1945 report *Science: The Endless Frontier*.[55] During the ensuing post-war era, a sharp delineation between "applied" and "basic" research[56] activities has become entrenched. The persistence of the basic-applied dichotomy is not merely conceptual but institutional. Funding regimes, evaluation criteria, and career incentives often reinforce categorical separation, even when research practice itself operates through hybrid, iterative dynamics. Basic research is often associated with curiosity-driven inquiry and long-term knowledge generation, while applied research is linked to problem-solving, innovation, and societal or commercial use. This distinction has served important purposes: protecting exploratory inquiry from short-term instrumental pressures and clarifying funding mandates. Moreover, it has allowed states to focus on supporting one or the other, depending on policy preferences: First, basic research during the Vannevar Bush-inspired post-war era; then innovation-oriented, applied research, from the 1990s onwards.[57]

---

[54] Palmås, K. (2024). Science theater on stage: Review of the play *The Right Way*, 2019-2020. *Science as Culture* 33(4), 579–587.

[55] Narayanamurti, V., Odumosu, T. & Vinsel, L. (2013). RIP: The Basic/Applied Research Dichotomy. *Issues in Science and Technology* 29(2).

[56] "Basic research" is also referred to as "fundamental research" or, sometimes, "pure research".

Yet historically, the boundary between basic and applied research has been fluid rather than fixed. Many transformative technologies have emerged from decades of fundamental investigation whose practical implications were initially unclear.[58] For example, advances in molecular biology, semiconductor physics, or materials science illustrate how conceptual breakthroughs and technological applications co-evolve. Conversely, attempts to solve concrete problems often generate new theoretical questions, reshaping fundamental understanding.

The case of lipid nanoparticle–enabled mRNA vaccines provides a striking illustration. Foundational work in RNA biology[59] and physical chemistry,[60] pursued without immediate clinical objectives, later proved indispensable in responding to the COVID-19 global health crisis. At the same time, attempts to translate molecular insights into therapeutic interventions generated new basic research questions. Rather than a linear progression from basic discovery to application, such trajectories reveal iterative feedback loops between experimentation, theory, and practical deployment.

Artificial intelligence research illustrates similar dynamics. Advances in machine learning often emerge from fundamental theoretical insights in statistics, mathematics, and computer science, yet their deployment in real-world applications rapidly reshapes research priorities and raises new scientific questions. The development of generative AI systems has also generated new challenges in model interpretability, alignment, and governance—issues that require not only technical innovation but also ethical, social, and institutional analysis.

Similar iterative dynamics can be observed in the history of semiconductor physics.[61] Foundational investigations into quantum behavior in solids, initially pursued without clear industrial targets, later enabled the transistor and integrated circuit. Subsequent engineering challenges in miniaturization and manufacturing, in turn, generated new theoretical questions in condensed matter physics. Here too, discovery and application evolved through reciprocal reinforcement rather than linear translation.

---

[57] Johnson, A. (2004). The end of pure science: science policy from Bayh-Dole to the NNI. In: Baird, D., Nordmann, A. & Schummer, J. (eds.) *Discovering the nanoscale*. Amsterdam; Washington, DC: IOS Press, 217–230.

[58] Stokes, D. E. (1997). *Pasteur's Quadrant: Basic Science and Technological Innovation*. Washington, DC Brookings Institution Press.

[59] Karlsson Hedestam, G. & Sandberg, R. (2026). Scientific background: Discoveries concerning nucleoside base modifications that enabled the development of effective mRNA vaccines against COVID-19 . Nobel Prize Outreach. https://www.nobelprize.org/prizes/medicine/2023/advanced-information/

[60] Horejs, C. (2021). From lipids to lipid nanoparticles to mRNA vaccines. *Nature Reviews Materials* 6, 1075–1076.

[61] Bernardo, C.P.C.V., Lameirinhas, R.A.M., de Melo Cunha, J.P. et al. (2024). A revision of the semiconductor theory from history to applications. *Discover Applied Sciences* 6, 316.

This dynamic suggests that the dichotomy between basic and applied research may obscure more than it clarifies, particularly in fields characterized by rapid technological transformation. For research excellence in groundbreaking technologies, the task is not to privilege basic over applied research, or vice versa, but to design institutional environments that sustain their long-term interaction and iterative feedback. Excellence, in this sense, lies in maintaining permeability across this boundary rather than reinforcing categorical separation.

## 5.2. Institutional structures reinforce a misleading polarization

Despite widespread recognition of their interdependence, institutional arrangements often reinforce a polarization between basic and applied research. Funding agencies may allocate resources through separate streams, evaluation panels may apply different criteria, and academic cultures may implicitly rank forms of inquiry according to perceived purity or utility.

Such arrangements can create unintended distortions. Research framed as "too applied" may struggle for recognition within systems oriented toward disciplinary novelty, while projects perceived as "too fundamental" may be deemed insufficiently responsive to societal challenges. In mixed panels, evaluative tensions can emerge when standards of success differ. These dynamics risk marginalizing research that operates productively in the space between conceptual exploration and practical engagement.

Moreover, the pressure to demonstrate short-term impact can incentivize premature claims of applicability, while defensive appeals to "basic research" may serve to shield projects from legitimate questions about broader relevance. The categories themselves can become rhetorical tools, invoked strategically rather than descriptively.

In the context of excellence initiatives focused on groundbreaking technologies, maintaining a rigid distinction between basic and applied research may hinder rather than support innovation. What is needed is not the erasure of conceptual clarity, but a reframing that recognizes technoscientific research as an iterative process in which discovery and application are co-constitutive.

Striking the right balance is consequential, where technically successful systems can generate enduring societal constraints when broader contexts are underexamined. For example, the early design of urban transportation infrastructures around private automobiles provides a longer-term example of technological lock-in.[62] Engineering excellence delivered unprecedented mobility, yet limited integration of urban planning, environmental foresight, and social equity considerations produced path dependencies that are costly to reverse. A contemporary illustration is the rapid deployment of AI-based facial recognition infrastructures[63]: technically groundbreaking systems were integrated into public and private environments before ethical, legal, and social considerations were fully addressed, resulting in forms of technological lock-in that are now difficult to reverse.[64] Likewise, emerging gene-drive technologies aimed at controlling mosquito populations demonstrate how powerful interventions can create irreversible ecological pathways if societal, environmental, and governance implications are not robustly assessed from the outset.[65]

---

[62] Urry, J. (2004). The 'System' of Automobility. *Theory, Culture & Society* 21(4-5), 25–39.
[63] Roussi, A. (2020). Resisting the rise of facial recognition. *Nature* 587(7834), 350–353.

Research infrastructures offer a concrete illustration of how this polarization can be overcome. Large-scale facilities, shared technological platforms, and distributed research networks routinely integrate fundamental inquiry, methodological innovation, and application-oriented development within a single organizational framework. They also embed governance questions—such as data stewardship, access policies, clinical translation, and public accountability—into their operational design. When structured inclusively, such infrastructures function as socio-technical environments in which scientific, technological, and humanistic expertise interact continuously rather than sequentially. Yet prevailing funding and evaluation systems often categorize infrastructures narrowly as either instruments of basic science or engines of innovation, thereby obscuring their integrative role and under-recognizing the contributions of the humanities and social sciences within them. Recognizing infrastructures as sites of iterative co-production highlights the limitations of rigid categorical distinctions and points toward institutional designs better aligned with the realities of contemporary technoscientific research.

## 5.3. Designing research systems for iterative co-production

Since research excellence in groundbreaking technologies depends on sustained interaction between fundamental inquiry and practical engagement, institutional design must reflect this reality. Rather than organizing research ecosystems around a binary distinction, systems should enable iterative feedback between conceptual exploration, technological development, and societal implementation. The following strategies outline how this can be achieved:

- **Structure funding schemes to support iterative trajectories.** Grant programs should allow projects to evolve across exploratory and translational phases without requiring artificial reclassification. Flexible funding mechanisms can enable research teams to pursue fundamental questions that emerge from applied challenges, and vice versa.

- **Align evaluation criteria with the "process" rather than the "category".** Assessment frameworks should focus on scientific rigor, originality, and long-term relevance rather than on whether a project is labeled basic or applied. Mixed evaluation panels should be trained to recognize the value of research that bridges conceptual and practical domains.

- **Protect exploratory inquiry while enabling engagement.** Safeguards for curiosity-driven research remain essential. At the same time, institutions should create pathways for engagement with industry, public agencies, and civil society that do not compromise academic autonomy. Clear governance structures can balance openness with integrity.

---

[64] Smith, M. & Miller, S. (2022). The ethical application of biometric facial recognition technology. *AI & Society* 37, 167–175.

[65] Fürer, C. L., Fischer, T. B., Suter, T., Winkler, M. S., & Knoblauch, A. M. (2026). Potential benefits, opportunities, risks and challenges of population suppression gene drive mosquitoes for malaria control described in the scholarly literature: a rapid scoping review. *Impact Assessment and Project Appraisal*, 1–23.

- **Recognize translational feedback as a source of fundamental insight.** Attempts to implement technologies in real-world contexts often reveal unanticipated phenomena, limitations, or ethical tensions. Research systems should treat such feedback not as peripheral troubleshooting but as opportunities for deepened theoretical understanding.

- **Avoid rhetorical polarization in research policy discourse.** Policymakers and institutional leaders should articulate excellence in terms that emphasize interdependence rather than opposition. Framing technoscientific research as inherently iterative can reduce unproductive competition between the formally different categories.

- **Encourage institutional spaces where basic and applied researchers as well as and humanities and social science researchers interact routinely.** Institutions should actively create shared environments in which fundamental, applied, and humanities and social science researchers interact on a routine basis rather than episodically. Research infrastructures are particularly well suited to this role. Far from being mere technical service providers, they function as socio-technical environments where epistemic norms, governance arrangements, collaboration patterns, and standards of evidence are established and reproduced. By bringing diverse forms of expertise into sustained proximity, infrastructures can reduce the artificial polarization between fundamental inquiry and practical application and instead support their iterative integration.

## Conclusion and Outlook

This perspective article emerged from an interdisciplinary workshop organized to explore how humanities and social sciences can be meaningfully integrated into technologically driven excellence environments, as part of the preparatory phase for the Swedish national excellence clusters in groundbreaking technologies.[66] Bringing together researchers from multiple natural sciences and engineering disciplines with philosophy, social sciences, science and technology studies, communication, and artistic disciplines, the workshop functioned as an effort to identify integration models, and articulate insights relevant for the continued development of excellence cluster framework. The conclusion and outlook presented here draw directly on these discussions, synthesizing their implications for how institutional arrangements must evolve if excellence initiatives centered on groundbreaking technologies are to generate not only technological breakthroughs but also socially robust and democratically legitimate forms of innovation.

The argument of this article is not that technological research lacks societal awareness, nor that the humanities and social sciences lack the tools to assist in research excellence for breakthrough technologies. Rather, it argues that the gap between these fields—in methods, incentive structures, and mutual understanding—is itself a risk factor for how emerging technologies will reshape societies. Closing this gap requires investment from both sides: technological research must make room for social inquiry early and substantively, and the humanities and social sciences must invest in long-term engagement with technological fields. Only in that way may scholars in the humanities and social science forge personal and professional networks, acquiring cross-disciplinary literacy.

---

[66] Swedish Research Council & Vinnova. (2026). *Excellence clusters for groundbreaking technologies*. https://www.vr.se/english/mandates/funding-and-promoting-research/excellence-clusters-for-groundbreaking-technologies.html

It follows that integration must be understood as bidirectional. While technological research benefits from humanistic and social inquiry, the humanities and social sciences may confront structural and epistemic challenges, adapting their conceptual frameworks to rapidly evolving technological systems, and strengthening methodological engagement with data-intensive environments. They may also find ways of positioning themselves within research ecosystems increasingly shaped by performance metrics and short-term innovation logics. Thus, the question is not whether integration is desirable, but whether the institutional conditions that make it possible will be created before the next generation of transformative technologies is already deployed. If not, we risk continuing down a path where technology increasingly creates problems rather than advantages for society.

Realizing this integration demands that it is structurally embedded rather than treated as compliance, outreach, or a late-stage add-on. Meaningful inclusion requires sustained collaboration, clearly defined roles, and long-term intellectual exchange within excellent research environments, with shared responsibility for shaping research roadmaps and criteria of success. Because many ethical and societal challenges are cross-cutting, such as responsibility, regulation, access, long-term impact, public trust, and governance recur across domains, comparative perspectives that connect work across multiple cutting edge technology development programs can reach deeper, more generalizable insights than any single technology domain can reliably generate in isolation. At the same time, autonomy must be combined with technical proximity: humanities and social sciences scholars should formulate research questions within their own disciplinary traditions and retain the critical distance needed to examine ethical, epistemic, legal, and societal dimensions, while maintaining close access to technological sites, data, and prototypes. Natural scientists and engineers can provide technical clarification and contextual understanding, and may serve as co-supervisors or technical advisors for cross-disciplinary doctoral and postdoctoral projects—without directing the framing of humanities and social sciences research.

Such integration also depends on deliberately cultivated social and intellectual spaces—recurring workshops, joint seminars, shared doctoral environments, and informal meeting arenas—that build mutual literacy, trust, and continuity over time. Broader interpretive practices, including collaboration with artistic disciplines such as photography and visual arts, can enrich how emerging technologies are represented, contextualized, and discussed, both intradisciplinary and with wider publics, strengthening civic understanding and legitimacy alongside scientific rigor. If emerging excellence initiatives in groundbreaking technologies embed these conditions—structural roles, cross-cluster learning, autonomous yet proximate collaboration, and sustained social infrastructure—then discovery and application can become genuinely co-constitutive. These arrangements must be established early in the research lifecycle, and not retrofitted after deployment. Only in that way can we avoid persistent lock-ins and downstream harms, and instead steer technological change toward long-term societal benefits.

The rapid rise of generative and agentic artificial intelligence makes this integration particularly urgent. As technological systems increasingly operate through language, cultural interpretation, and normative alignment, the competences traditionally cultivated in the humanities and social sciences become integral to the design, governance, and responsible deployment of emerging technologies. Integrating these perspectives early in research environments will therefore be essential for ensuring that future technological breakthroughs remain socially intelligible, ethically grounded, and democratically legitimate.

## Acknowledgements

This work was supported by the Swedish Research Council through projects #2025-07534 ("Quantitative methods to measure and predict protein folding and misfolding with single-molecule resolution") and #2025-07556 ("Quantitative single-molecule microscopy to advance biomedicine"), which funded the activities leading to this perspective. The authors acknowledge the contributions of all participants in the roundtable discussions held at Ågrenska Villan, University of Gothenburg, on 30 January 2026, whose insights and critical reflections informed the development of the perspectives presented in this article. The workshop brought together researchers affiliated with additional Swedish Research Council–funded projects, including #2025-07595 ("The Culture Code: Deep Sustainable AI"), #2025-07479 ("National excellence cluster in AI-driven antibiotic innovation"), #2025-07465 ("An excellence centre for RNA-based precision therapies"), #2025-07507 ("Adaptive soft material systems for technological innovations"), and #2025-07583 ("Innovations in Ultrafast Laser Science - Source Development and Applications"), whose participation enriched the interdisciplinary dialogue. We thank Christian Munthe (University of Gothenburg), Maria Tenje (University of Uppsala), and Nina Wormbs (KTH Royal Institute of Technology) for insightful contributions during the roundtable discussions. We also thank Agnese Callegari from the Soft Matter Lab at the University of Gothenburg and Aykut Argun from Vintergatan Photography for their valuable support.